# Toward Emerging Topic Detection for Business Intelligence: Predictive Analysis of 'Meme' Dynamics

**Kristin Glass**[1]     **Richard Colbaugh**[1,2]

[1] New Mexico Institute of Mining and Technology, Socorro, NM USA
[2] Sandia National Laboratories, Albuquerque, NM USA

**Abstract**

Detecting and characterizing emerging topics of discussion and consumer trends through analysis of Internet data is of great interest to businesses. This paper considers the problem of monitoring the Web to spot emerging *memes* – distinctive phrases which act as "tracers" for topics – as a means of early detection of new topics and trends. We present a novel methodology for predicting which memes will propagate widely, appearing in hundreds or thousands of blog posts, and which will not, thereby enabling discovery of *significant* topics. We begin by identifying measurables which should be predictive of meme success. Interestingly, these metrics are not those traditionally used for such prediction but instead are subtle measures of meme dynamics. These metrics form the basis for learning a classifier which predicts, for a given meme, whether or not it will propagate widely. The utility of the prediction methodology is demonstrated through analysis of memes that emerged online during the second half of 2008.

## 1. Introduction

The enormous popularity of "social media", such as blogs, forums, and social networking sites, represents a significant challenge to standard business models and practices, as these media move the control of information from companies to consumers [e.g. 1-5]. However, social media also offer an unprecedented opportunity to increase business responsiveness and agility. For example, recent surveys reveal that 32% of the nearly 250 million bloggers worldwide regularly give opinions on products and brands, 71% of active Internet users read blogs, and 70% of consumers trust opinions posted online by other consumers [6,7]. Thus social media is a vast source of business-relevant opinions. Moreover, this information has a reach that rivals any traditional media and an influence which substantially exceeds standard advertising channels.

Businesses are therefore strongly motivated to pay attention to social media and other online information sources. For instance, it is crucially important for companies to be able to detect emerging topics of discussion and consumer trends as soon as possible. Customer complaints and other negative information are much easier to address if discovered quickly, while early positive "buzz" can be leveraged and amplified. Identifying nascent consumer interest in subjects and trends which are relevant to company business can be of great strategic advantage. Indeed, the relevance and timeliness of the information available in social media has the potential to revolutionize the way business is conducted in many sectors.

A central challenge in leveraging the information present in social media is the enormous scale of the problem. The data of interest to a particular business are buried in the vast, and largely irrelevant, output of millions of bloggers and other online content producers. Consequently, effectively exploiting these data requires development of new, automated methods of analysis [1-5].

This paper considers the problem of detecting and characterizing emerging topics and trends in social media. Automated topic discovery is of great interest in research fields ranging from business to computer science and application domains such as marketing, finance, human health, and national security; see, e.g., [1-3] for business-oriented reviews. Recently [8] proposed that monitoring social media to spot emerging *memes* – distinctive phrases which propagate relatively unchanged online and act as "tracers" for discrete cultural units – can enable early discovery of new topics and trends. These researchers present an elegant solution to the meme detection problem and show that their algorithm is efficient enough to allow Web-scale analysis. However, a challenge with the meme-tracking method developed in [8] is the fact that the vast majority of online memes attract little attention before fading into obscurity. In contrast, in most business applications we are interested in those memes, and the underlying topics, that reach a nontrivial fraction of the population.

This consideration motivates our interest in *predictive* analysis of meme dynamics: we wish to identify those memes which will go on to attract significant attention, and to do so very early in the meme lifecycle. Observe that such a capability is essential for practical emerging topic discovery, as it would enable early detection of the emer-gence of *significant* topics and trends. Most approaches to predictive analysis of social diffusion phenomena like this assume, either explicitly or implicitly, that diffusion events which "go viral" are qualitatively different from those that don't, and devote attention on trying to identify these crucial differences (see [9,10] for background on social diffusion). Recent research calls into question this basic premise and, indeed, indicates there may be fundamental limits to what can be predicted about social dynamics. For example, the studies reported in [11-14] indicate that the "intrinsic" attributes commonly believed to be important when assessing the likelihood of adoption of popular cultural products, such as the quality of the product itself, do not possess much predictive power. This research offers evidence that, when individuals are influenced by the actions of others, it may not be possible to obtain reliable predictions using methods which focus on intrinsics alone; instead, it may be necessary to incorporate aspects of social influence into the prediction process.

Recognizing this challenge, this paper proposes that generating useful predictions about social dynamics requires careful consideration of the way individuals influence one another through their social networks. We presents a powerful new methodology for predictive analysis of social diffusion which exploits information about network topology and dynamics to accurately forecast which memes will propagate widely, appearing in hundreds or thousands of blog posts, and which will not. The particular network features used by the prediction algorithm are those identified as likely to be predictive of meme success by our recently proposed predictability assessment method [11]. Interestingly, the metrics nominated by this theoretical analysis as the ones expected to possess significant predictive power turn out to be fairly subtle measures of the network dynamics associated with early meme diffusion. Meme prediction is accomplished by learning an algorithm which, based upon very early network dynamics, is able to successfully identify which memes will ultimately propagate widely and which will not.

The paper makes three main contributions. First, we propose a set of novel network dynamics-based metrics which possess significant predictive power for social diffusion processes like meme propagation; indeed, these metrics are found to be considerably more predictive than standard measures. Second, we develop a machine learning-based classification algorithm which employs these network dynamics metrics to accurately predict, very early in the lifecycle of a meme of interest, whether that meme will diffuse widely or not. Third, the utility of the prediction algorithm, and the power of network-based predictive metrics, are demonstrated through an empirical study involving "successful" and "unsuccessful" memes associated with topics of discussion that emerged in social media during the second half of 2008. Perhaps surprisingly, we find that although memes typically propagate for weeks, useful predictions can be made *within the first twelve hours* after a meme is detected.

## 2. Problem Formulation

This section begins by defining the class of memes of interest and providing a more precise statement of the meme prediction problem. We then introduce the datasets we use to investigate meme prediction and briefly characterize relevant attributes of the dynamics of meme propagation.

The goal of this paper is to develop a methodology for understanding and predicting the way memes – distinctive phrases which act as "tracers" for topics – diffuse through online news and social media. We focus on phrases which 1.) appear in online sources as quoted material, and 2.) propagate largely unchanged through sequences of news articles and blog posts. Our main source of data on meme dynamics is the publicly available datasets archived at http://memetracker.org [15] by the authors of [8]. Briefly, the study [8] develops an efficient and elegant algorithm for meme detection and applies the algorithm to data obtained from Spinn3r (http://spinn3r.com). The raw Spinn3r data used in [8] consist of all news articles and blog posts published on the approximately 20 000 news sites and 1.6 million blogs indexed by Spinn3r during the period 1 August to 31 December 2008. Application of the meme detection method proposed in [8] to the Spinn3r data resulted in the extraction of over 112 million phrases, and a significant portion of these data have been made available at [15].

We are interested in developing capabilities to perform two main operations: 1.) meme *monitoring*, involving identification of a few good "sensor blogs" through which to observe the emergence/evolution of memes, and 2.) meme *prediction*, the goal of which is to distinguish successful and unsuccessful memes early in their lifecycle. More precisely, we have the following predictive analysis problems:

- Monitoring: 1.) investigate whether there exist good *sensor blogs* for memes, that is, a small number of blogs that consistently detect successful memes early; if so, 2.) characterize these sensor blogs and employ them to increase the efficiency and effectiveness of the meme detection method of [8].

- Prediction: 1.) identify measurables which are predictive of meme success (e.g., post sentiment, early meme dynamics), and 2.) use these predictive measurables as the basis for classifying memes into two groups – those which will ultimately be successful (here, acquire ≥1000 posts) and those that will be unsuccessful (attract ≤100 posts) – very early in the meme lifecycle.

To support an empirical evaluation of our proposed solutions to the above problems, we downloaded from [15] the time series data for slightly more than 70 000 memes. These data contain, for each meme M, a sequence of pairs $(t_1, URL_1)_M, (t_2, URL_2)_M, \ldots, (t_T, URL_T)_M$, where $t_k$ is the time of appearance of the kth blog post or news article that contains at least one mention of meme M, $URL_k$ is the URL of the blog or news site on which that post/article was published, and T is the total number of posts that mention meme M. From this set of time series we randomly se-lected 100 "successful" meme trajectories, defined as those corresponding to memes which attracted at least 1000 posts during their lifetimes, and 100 "unsuccessful" meme trajectories, defined as those whose memes acquired no more than 100 total posts. It is worth noting that, in assembling the data in [15], all memes which received fewer than 15 total posts were deleted, and that ~50% of the remaining memes have <50 posts; thus the large majority of memes are unsuccessful by our definition (as well as according to the criteria of most business applications). Figure 1 depicts the distribution of meme success for the full set of ~70 000 memes obtained from [15].

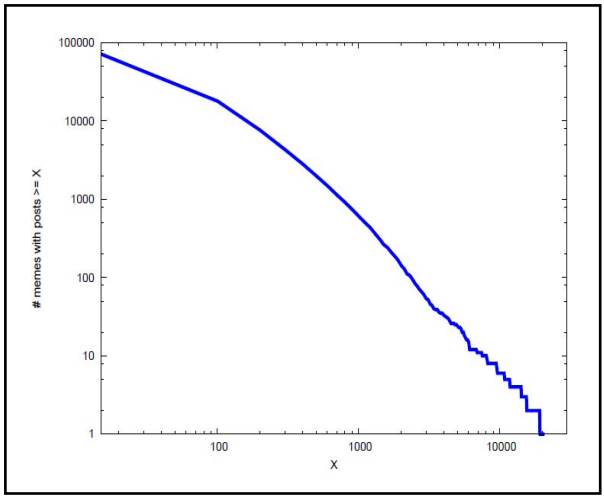

**Figure 1.** Cumulative distribution of meme success. The log-log plot shows the number of memes (vertical axis) which acquired at least a given number of posts (horizontal axis) during their lifetimes.

Two other forms of data were collected for this study: 1.) a large Web graph which includes websites (URLs) that appear in the meme time series, and 2.) samples of the text surrounding the memes in the posts which contain them. More specifically, we sampled the URLs appearing in the time series for our set of 200 successful and unsuccessful memes and performed a Web crawl that employed these URLs as "seeds". This procedure generated a Web graph, denoted $G_{web}$, that consists of approximately 550 000 vertices/websites and 1.4 million edges/hyperlinks, and includes essentially all of the websites which appear in the meme time series. To obtain samples of text surrounding memes in posts, we randomly selected ten posts for each meme and then extracted from each post the paragraph which contains the first mention of the meme.

Meme dynamics possesses several characteristics which are likely to make predictive analysis challenging. For example, as shown in Figure 1, the distribution for meme success is strongly right-skewed, with most memes receiving relatively little attention and a few attracting significant interest. This property may be a reflection of the role of social influence in meme dynamics: individuals and organizations become aware of memes and judge their appeal in part by observing others, and as a consequence successful memes may attract attention not because they are particularly interesting but instead simply because they've attracted attention in the past [e.g., 12,14]. It is known that predicting the evolution of such "rich get richer" phenomena using standard methods is quite difficult [11-14].

Memes also exhibit highly variable times to acquire their first few posts and to accumulate their final tally of posts. Figure 2 reports the mean and median times required for successful and unsuccessful memes to attract five, ten, and their total number of posts. It is interesting to note that the median times for unsuccessful posts to attract early posts is actually *shorter* than the corresponding times for successful memes. This figure also provides an illustration, taken from [8], of the evolution of several successful memes. It can be seen that some of these memes receive their posts quickly while others become prominent only after significant time has elapsed.

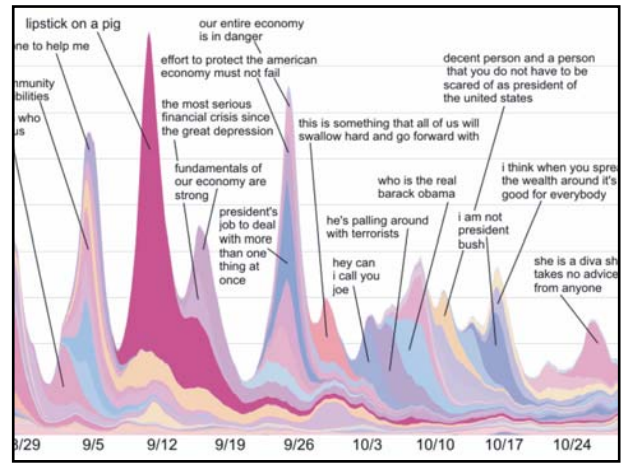

| Early Meme Dynamics | | |
|---|---|---|
| **Successful Memes (>1000 posts)** | | |
| #posts | mean (hr) | median (hr) |
| 5 | 108 | 18.5 |
| 10 | 171 | 41.5 |
| total | ~4400 | ~4400 |
| **Unsuccessful Memes (<100 posts)** | | |
| #posts | mean (hr) | median (hr) |
| 5 | 375 | 10.1 |
| 10 | 765 | 30.5 |
| total | ~1010 | ~410 |

**Figure 2.** Meme dynamics. In the "stacked" plot at top, thread thickness corresponds to number of posts/articles mentioning the particular meme during that time period (horizontal axis) [8]. The table at bottom reports the mean and median time (in hours) required for successful and unsuccessful memes to acquire five posts, ten posts, and their total number of posts.

## 3 Predictive Analysis

In this section we begin by summarizing the application of the predictability assessment process [11,12] to a simple model of meme diffusion. This procedure identifies two features of the network dynamics which should be useful for distinguishing successful and unsuccessful memes early in their lifecycle. We then consider the meme monitoring problem, in which the objective is to identify and characterize blogs and other online sources that consistently detect successful memes early in their lifecycles. Discovering such news sources would enable more efficient Web monitoring and is thus of considerable direct interest. These sources may also be useful for meme prediction, as the appearance of a meme on one or more of these sites may be an exploitable early indicator of meme success. Finally we address meme prediction, and develop a machine learning-based classification algorithm which employs novel network dynamics metrics to accurately predict, very early in a meme's lifecycle, whether that meme will propagate widely or not. The performance of the prediction algorithm is illustrated through an empirical study involving successful and unsuccessful memes associated with topics of discussion that emerged in social media during 2008.

### 3.1 Predictability assessment

Here we briefly summarize the results of applying the predictability assessment procedure [11,12] to the task of identifying measurables which should be predictive of meme success. The discussion begins with brief, intuitive reviews of the predictability assessment process and a general framework for modeling social diffusion, and then describes the results of applying this analytic process to meme dynamics. Readers interested in a comprehensive mathematical presentation of predictability assessment and social diffusion modeling are referred to [11,12].

**Predictability.** Consider a simple model of information diffusion, in which individuals combine their own beliefs and opinions regarding a new piece of information with their observations of the actions of others to arrive at their decisions about whether to pass along the information. In such situations it can be quite difficult to determine which characteristics of the diffusion process, if any, are predictive of things like the speed or ultimate reach of the diffusion [9-12]. In [11,12] we propose a mathematically rigorous approach to predictability assessment which, among other things, permits identification of features of social dynamics which should have predictive power; we now summarize this assessment methodology.

The basic idea behind the proposed approach to predictability analysis is simple and natural: we assess predictability by answering questions about the reachability of diffusion events. To obtain a mathematical formulation of this strategy, the behavior about which predictions are to be made is used to define the system *state space subsets of interest* (SSI), while the particular set of candidate measurables under consideration allows identification of the *candidate starting set* (CSS), that is, the set of states and system parameter values which represent initializations that are consistent with, and equivalent under, the presumed observational capability. As a simple example, consider an online market with two products, A and B, and suppose the system state variables consist of the current market share for A, ms(A), and the rate of change of this market share, r(A) (ms(B) and r(B) are not independent state variables because ms(A) + ms(B) = 1 and r(A) + r(B) = 0); let the parameters be the advertising budgets for the products, bud(A) and bud(B). The producer of item A might find it useful to define the SSI to reflect market share dominance by A, that is, the subset of the two-dimensional state space where ms(A) exceeds a specified threshold (and r(A) can take any value). If only market share and advertising budgets can be measured then the CSS is the one-dimensional subset of state-parameter space consisting of the initial magnitudes for ms(A), bud(A), and bud(B), with r(A) unspecified (the one-dimensional "uncertainty" in the CSS reflects the fact that r(A) is not measurable).

Roughly speaking, the approach to predictability assessment proposed in [11,12] involves determining how probable it is to reach the SSI from a CSS and deciding if these reachability properties are compatible with the prediction goals. If a system's reachability characteristics are incompatible with the given prediction question – if, say, "hit" and "flop" in the online market example are both likely to be reached from the CSS – then the situation is deemed unpredictable. This setup permits the identification of candidate predictive measurables: these are the measurable states and/or parameters for which predictability is most sensitive [11,12]. Continuing with the online market example, if trajectories with positive early market share rates r(A) are much more likely to yield market share dominance for A than are trajectories with negative early r(A), then the situation is unpredictable (because the outcome depends sensitively on r(A) and this quantity is not measured). Moreover, this analysis suggests that market share rate is likely to possess predictive power, so it may be possible to increase predictability by adding the capacity to measure this quantity.

**Model.** In social diffusion, people are affected by what others do. This is easy to visualize in the case of disease transmission, with infections being passed from person to person. Information, such as that in the topics of discussion underlying memes, can also propagate through a population, as individuals become aware of information and persuaded of its relevance through their social and information networks. The dynamics of information diffusion can therefore depend upon the topological features of the pertinent networks, for instance the presence of highly connected blogs in a social media network (see, e.g., [10]). This dependence suggests that, in order to understand the predictability of social diffusion phenomena and in particular to identify features which possess predictive power, it is necessary to perform predictability assessment on social and information networks with realistic topologies.

Specifically, the social diffusion models examined in this study possess networks with four topological properties that are ubiquitous in real-world social and information networks and have the potential to impact diffusion dynamics:

- *right-skewed degree* – the property that most vertices have only a few network neighbors while a few vertices have a great many neighbors;
- *transitivity* – the property that the network neighbors of a given vertex have a heightened probability of being connected to one another;
- *community structure* – the presence of densely connected groupings of vertices which have only relatively few links to other groups;
- *core-periphery structure* – the presence of a small group of "core" vertices which are densely connected to each other and are also close to the other vertices in the network.

It is shown in [11] that *stochastic hybrid dynamical systems* (S-HDS) provide a powerful mathematical formalism with which to represent social diffusion on realistic networks. An S-HDS is a feedback interconnection of a discrete-state stochastic process, such as a Markov chain, with a family of continuous-state stochastic dynamical systems [11]. Combining discrete and continuous dynamics within a unified, computationally tractable framework offers an expressive, scalable modeling environment that is amenable to formal analysis. In particular, S-HDS models can be used to represent social diffusion on large-scale networks with right-skewed degree distributions, transitivity, community structure, and core-periphery structure [11,16].

As an intuitive illustration of the way S-HDS enable effective, tractable modeling of complex network phenomena, consider the task of modeling diffusion on a network that possesses community structure. As shown in Figure 3, this diffusion consists of two components: 1.) *intra-community dynamics*, involving frequent interactions between individuals within the same community and the resulting gradual change in the concentrations of "infected" (red) individuals, and 2.) *inter-community dynamics*, in which the "infection" jumps from one community to another in an abrupt manner. S-HDS models are a natural framework for representing these dynamics, with the S-HDS continuous system modeling the intra-community dynamics (e.g., via stochastic differential equations), the discrete system capturing the inter-community dynamics (e.g., using a Markov chain), and the interplay between these dynamics being represented by the S-HDS feedback structure. A detailed description of the manner in which S-HDS models can be used to capture information dynamics on networks that possess each of the topological structures of interest, and their combination, is given in [16].

**Results.** We have applied the predictability assessment methodology summarized above to the meme prediction problem, and we now summarize the main conclusions of this study. A complete discussion of this investigation, including source code and all analytic results, is given in [16]. The analysis uses the mathematically rigorous predictability assessment procedure summarized above, in combination with empirically-grounded S-HDS models for social dynamics, to characterize the predictability of social diffusion on networks with realistic degree distributions, transitivity, community structure, and core-periphery structure. The main finding of the study, from the perspective of this paper, is a demonstration that predictability of these diffusion models depends crucially upon social and information network topology, and in particular on a network's community and core-periphery structures.

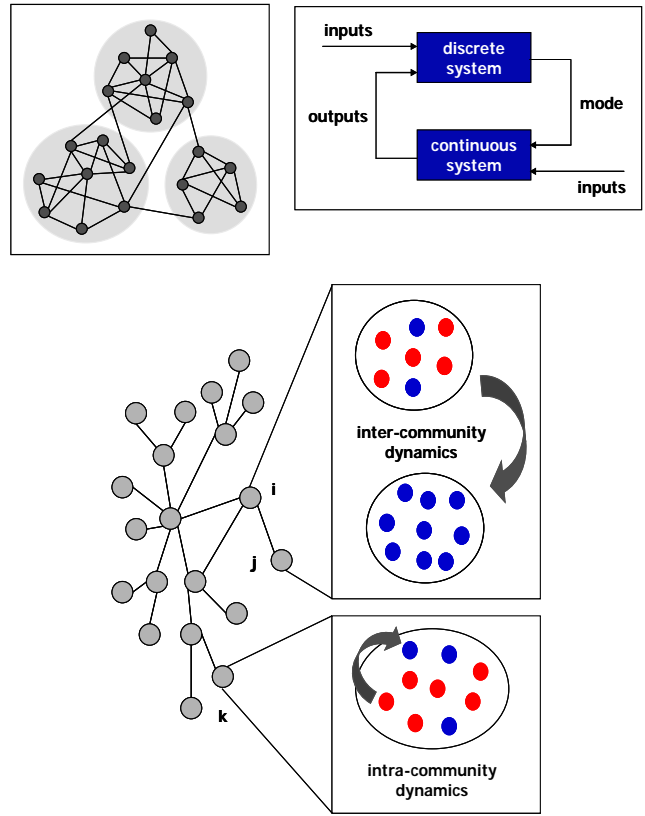

**Figure 3.** Modeling diffusion on networks with community structure via S-HDS. The cartoon at top left depicts a network with three communities. The cartoon at bottom illustrates diffusion *within* a community k and *between* communities i and j. The schematic at top right shows the basic S-HDS feedback structure; the discrete and continuous systems in this framework model the inter-community and intra-community diffusion dynamics, respectively.

In order to describe these theoretical results more quantitatively and leverage them for prediction, it is necessary to specify mathematical definitions for network communities and core-periphery structure. Community structure is widely recognized to be important in real-world networks, and there exists a range of qualitative and quantitative defini-tions for this concept. Here we adopt the modularity-based definition proposed in [17], whereby a good partitioning of a network's vertices into communities is one for which the number of edges between putative communities is smaller than would be expected in a random partitioning. To be concrete, a modularity-based partitioning of a network into two communities maximizes the modularity Q, defined as

$$Q = s^T B s / 4m,$$

where m is the total number of edges in the network, the partition is specified with the elements of vector s by setting $s_i = 1$ if vertex i belongs to community 1 and $s_i = -1$ if it belongs to community 2, and the matrix B has elements $B_{ij} = A_{ij} - k_i k_j / 2m$, with $A_{ij}$ and $k_i$ denoting the network adjacency matrix and degree of vertex i, respectively. Partitions of the network into more than two communities can be constructed recursively [17]. Note that modularity-based community partitions can be efficiently computed for large social networks and effectively implemented even with incomplete network topology data [16].

With this definition in hand, we are in a position to present the first candidate predictive feature nominated by the theoretical predictability assessment [16]: early dispersion of a diffusion process across network communities should be a reliable predictor that the ultimate reach of the diffusion will be significant; in particular, this measure should be more predictive than the early volume of diffusion activity. The basic idea is illustrated in Figure 4.

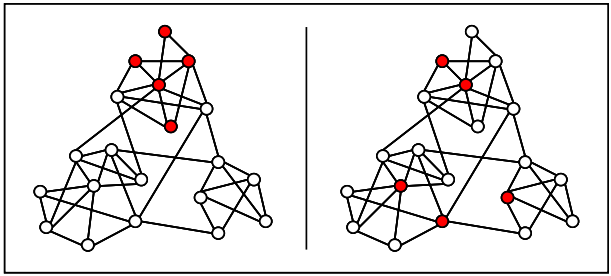

**Figure 4.** Early dispersion across communities is predictive. The cartoon illustrates the predictive feature associated with community structure: social diffusion initiated with five "seed" individuals is much more likely to propagate widely if these seeds are dispersed across three communities (right) rather than concentrated within a single community (left). Note that in [16] this result is established for networks of realistic scale and not simply for "toy" networks.

Analogously to the situation with network communities, there exist a wide range of qualitative and quantitative descriptions of the core-periphery structure found in real-world networks. Here we adopt the characterization of network core-periphery which results from *k-shell decomposition*, a well-established technique in graph theory which is summarized, for instance, in [18]. To partition a network into its k-shells, one first removes all vertices with degree one, repeating this step if necessary until all remaining vertices have degree two or higher; the removed vertices constitute the 1-shell. Continuing in the same way, all vertices with degree two (or less) are recursively removed, creating the 2-shell. This process is repeated until all vertices have been assigned to a k-shell, and the shell with the highest index, the $k_{max}$-shell, is the core of the network.

Given this definition, we are in a position to report the second candidate predictive feature nominated by the theoretical predictability assessment [16]: early diffusion activity within the network $k_{max}$-shell should be a reliable predictor that the ultimate reach of the diffusion will be significant; in particular, this measure should be more predictive than the early volume of diffusion activity. An intuitive illustration of this result is depicted in Figure 5.

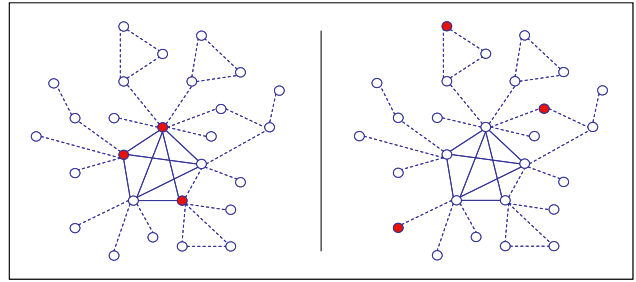

**Figure 5.** Early diffusion within the core is predictive. The cartoon illustrates the predictive feature associated with k-shell structure: social diffusion initiated with three "seed" individuals is much more likely to propagate widely if these seeds reside within the network's core (left) rather than at its periphery (right). Note that in [16] this result is established for networks of realistic scale and not simply for "toy" networks.

### 3.2 Monitoring

It is of great interest to determine whether there exist blogs and other online sources which are "good sensors" for emerging topics and memes, that is, are capable of detecting successful memes "early" in their lifecycles. For the purposes of this study, "early" is defined to be within the first 3% of the total duration of a meme's lifespan; qualitatively similar results are obtained using other definitions for early (e.g., first 5% of lifespan, first 3% or 5% of total accumulated blog posts). We adopt a very simple approach to this problem: we seek blogs which are "better than random" at early detection of successful memes. Other approaches are possible, of course, such as attempting to identify the set of blogs which enables optimally fast detection of new memes subject to a fixed "budget" on the number of blogs that can be monitored [16]. However, the present formulation is useful for understanding the properties and roles of individual blogs, for instance to allow influential information sources to be determined, and so is perhaps of more fundamental interest.

Consider the following methodology for identifying "better than random" sensor blogs:

1. Assemble a set of 50 successful memes, that is, memes which acquire at least 1000 posts during their lifetimes, from the set of memes archived at [15]. These memes are randomly selected from the set that remains in the dataset after removal of the 200 memes described above.

2. Form the null hypothesis:
   - blogs post on topics of interest, and they are equally good at posting early (so each blog may post on any topic, and thus mention any meme, but no blog is inherently superior at posting early);
   - the probability that blog B will successfully post early on a new meme is independent of B's previous performance on other memes (so Prob{blog has n successes} follows a binomial distribution).

3. Determine whether any blogs are good *early sensor* (ES) blogs, i.e., post early on more of the successful memes than would be expected under the null hypothesis.

4. Characterize the graph-topological properties of any such ES blogs.

Analysis of the time series data for the set of 50 successful memes considered here reveals that, of the 1.6 million blogs and 20 000 news feeds in the raw Spinn3r data [15], approximately 2400 online sources post early for at least one meme. Of these sources, only 33 post early at a rate which is statistically significantly better than would be expected under the null hypothesis ($p < 0.05$). This set of ES blogs is listed in Figure 6.

A few of the sources listed in Figure 6 perform much better than random, detecting more than half of the 50 successful memes within the first 3% of their lifespan. Interestingly, four of the ES blogs are also good at *avoiding* making posts which mention unsuccessful memes. More precisely, the four ES blogs shown in bold type in Figure 6 are less likely than random ($p < 0.05$) to mention memes which attract fewer than 25 total posts. Because these four information sources are simultaneously more likely than random to mention successful memes early and less likely than random to mention unsuccessful memes at any time, it is plausible that these sources are *influential*, and actually contribute to the success of memes by mentioning them. It is worth noting that, as shown in [16], knowledge of the identities of ES blogs can be used to improve the efficiency of the meme detection algorithm proposed in [8] without sacrificing meme coverage.

It would be very useful to identify easy-to-measure features of ES blogs to enable them to be efficiently and conveniently recognized (e.g., to allow discovery of good early sensors in a new domain). Simple graph topology measures, such as in-degree and closeness centrality, are natural candidate features and can explain some of the success of ES blogs. For instance, blog in-degree is correlated with the number of memes detected early by a blog. Unfortunately, though, this relationship is not sufficiently strong to be useful in discovering ES blogs.

---

**Early Sensors**

| | |
|---|---|
| us.rd.yahoo.com | ap.google.com |
| startribune.com | chicago.craigslist.org |
| cnn.com | chicagotribune.com |
| c.moreover.com | huffingtonpost.com |
| uk.news.yahoo.com | news.google.com |
| leftword.blogdig.net | online.wsj.com |
| news.originalsignal.com | buzz.originalsignal.com |
| washingtonpost.com | cnn ... .wordpress.com |
| blogs.abcnews.com | **elections.foxnews.com** |
| boston.com | **floppingaces.net** |
| x.techwheat.com | forums.hannity.com |
| breitbart.com | gretawire.foxnews.com |
| guardian.co.uk | news.smh.com.au |
| instablogs.com | **talkleft.com** |
| nydailynews.com | **thinkprogress.org** |
| salon.com | voices.washingtonpost.com |
| scienceblogs.com | |

---

**Figure 6.** "Early sensor" blogs and online news sources. This set of 33 online sources is unusually good at detecting successful memes early in their lifecycles. The four sources in bold type are also unusually good at avoiding the mention of unsuccessful memes.

Recall that our predictability analysis indicates that vertex k-core value may be predictive of meme success, which suggests the possibility that blogs with high k-core values may be good early sensors. Analysis of the Web graph $G_{web}$ associated with the meme data offers support for this hypothesis. In particular, we find:

- 67% of the ES blogs (listed in Figure 6) are located in the "core" of $G_{web}$, defined as the top 0.1% of websites in the graph as ranked by k-core index value;

- 64% of "strong" ES blogs (those which detect at least 25% of successful memes early) are located in the 324-vertex $k_{max}$-shell; note that, in comparison, the set of the top 324 websites in $G_{web}$ as ranked by in-degree contains 36% of strong ES blogs.

These results, while preliminary, suggest that k-shell decomposition provides a useful way to identify blogs and other online sources to monitor to enable emerging topics, and their associated memes, to be detected early in their lifecycles. For example, in the case of the memes investigated here, the $k_{max}$-shell represents a small set of websites to monitor, is computationally easy to discover, and contains most of the strong ES blogs identified through analysis of meme dynamics.

### 3.2 Prediction

We now turn to the task of developing a machine learning-based classifier which is capable of accurately predicting, very early in the lifecycle of a meme of interest, whether that meme will propagate widely. Two types of classification algorithm were tested, one simple (standard naïve Bayes [19]) and one sophisticated (the Avatar ensembles of decision trees algorithm [20]), to allow the robustness of the proposed approach to meme prediction to be evaluated. While the two classifiers produce qualitatively similar results, the Avatar algorithm, denoted A-EDT, is substantially more accurate; thus in what follows only the results obtained with A-EDT are reported.

Recall that the task of interest is to learn a classifier which takes as inputs some combination of relevant post content and meme dynamics and accurately predicts whether a given meme will ultimately be successful (acquire ≥1000 posts during its lifetime) or unsuccessful (attract ≤100 total posts). We employ standard ten-fold cross-validation to estimate the accuracy of our classifier. More specifically, the set of 200 memes (100 successful and 100 unsuccessful) is randomly partitioned into ten subsets of equal size, and the A-EDT algorithm is successively "trained" on nine of the subsets and "tested" on the held-out subset in such a way that each of the ten subsets is used as the test set exactly once.

A crucial aspect of the analysis is determining which characteristics of memes and their dynamics, if any, possess exploitable predictive power. We consider three classes of features:

- *language-based* measures, such as the sentiment and emotion expressed in the text surrounding memes in posts;
- *simple dynamics-based* metrics, capturing the early volume of posts mentioning the meme of interest and the rate at which this volume is increasing;
- *network dynamics-based* features, such as those identified through the predictability analysis summarized in Section 3.1.

We now describe each of these feature classes. Consider first the language-based measures. Each "document" of text surrounding a meme in a posts is represented by a simple "bag of words" feature vector $x \in \Re^{|V|}$, where the entries of x are the frequencies with which the words in the vocabulary set V appear in the document. The sentiment and emotion of a document may be quantified very simply through the use of appropriate lexicons. Let $s \in \Re^{|V|}$ denote a lexicon vector, in which each entry of s is a numerical "score" quantifying the sentiment/emotion intensity of the corresponding word in the vocabulary V. The aggregate sentiment/emotion of document x can be computed as

$$\text{score}(x) = s^T x / s^T 1,$$

where 1 is a vector of ones. Thus score(.) estimates the sentiment or emotion of a document as a weighted average of the sentiment or emotion scores for the words comprising the document. (Note that if no sentiment or emotion information is available for a particular word in V then the corresponding entry of s is set to zero.)

To characterize the emotion content of a document we use the Affective Norms for English Words (ANEW) lexicon, which consists of 1034 words that were assigned numerical scores with respect to three emotional "axes" – happiness, arousal, and dominance – by human subjects [21]. Previous work had identified this set of words to bear meaningful emotional content [21]. Positive or negative sentiment is quantified by employing the "IBM lexicon", a collection of 2968 words that were assigned {positive, negative} sentiment labels by human subjects [22]. Observe that this simple approach generates four language features for each meme: the happiness, arousal, dominance, and positive/negative sentiment of the text surrounding that meme in the post containing it. As a preliminary test, we computed the mean emotion and sentiment of content surrounding the 100 successful and 100 unsuccessful memes in our dataset. On average the text surrounding successful memes is happier, more active, more dominant, and more positive than that surrounding unsuccessful memes, and this difference is statistically significant (p<0.0001). Thus it is at least plausible that these four language features may possess some predictive power regarding meme success.

Consider next two simple dynamics-based features, defined to capture the basic characteristics of the early evolution of meme post volume:

- #posts($\tau$) – the cumulative number of posts mentioning the given meme by time $\tau$ (where $\tau$ is small relative to the typical lifespan of memes);
- post rate($\tau$) – a simple estimate of the rate of accumulation of such posts at time $\tau$.

Here we adopt a simple finite difference definition for post rate given by post rate($\tau$) = (#posts($\tau$) – #posts($\tau/2$)) / ($\tau/2$); of course, more robust rate estimates could be used.

The simple dynamics-based measures of early meme diffusion defined above, while potentially useful, do not characterize the manner in which a meme propagates over the underlying social or information networks. Recall that the predictability assessment summarized in Section 3.1 suggests that both early dispersion of diffusion activity across network communities and early diffusion activity within the network core ought to be predictive of meme success. Additionally, the empirical analysis identifying good early sensor blogs presented in Section 3.2 indicates that another network feature, the ES blogs in the Web graph, may also possess predictive power. The insights offered by these theoretical and empirical analyses motivate the definition of three network dynamics-based features for meme prediction:

- community dispersion($\tau$) – the cumulative number of network communities in the Web graph $G_{web}$ that contain at least one post mentioning the meme by time $\tau$;
- #k-core blogs($\tau$) – the cumulative number of blogs in the $k_{max}$-shell of Web graph $G_{web}$ that contain at least one post mentioning the meme by time $\tau$;

- #ES blogs($\tau$) – the cumulative number of ES blogs that contain at least one post mentioning the meme by time $\tau$.

The quantities specified in the first two definitions can be efficiently computed using fast algorithms for partitioning a graph into its communities and for identifying the graph's $k_{max}$-shell (code implementing these algorithms is available in [16]). Consequently, these features are readily computable for very large graphs. The third feature, #ES blogs($\tau$), is trivial to compute given the identities of the ES blogs listed in Figure 6.

We now summarize the main results of the prediction study (see [16] for a more complete description of the results). First, implementing the A-EDT algorithm with the four language features for the task of predicting which memes will be successful yields a prediction accuracy of 66.5% (ten-fold cross-validation). Since simply guessing "successful" for all memes gives an accuracy of 50%, it can be seen that these simple language "intrinsics" are not very predictive. For what it's worth, the ANEW score for "arousal" and the IBM measure of sentiment are the most predictive language features (though of course neither is very predictive).

In contrast, this investigation reveals that the features characterizing early network dynamics of memes do possess predictive power, and in fact are useful even if only very limited early time series is available for use in prediction. More quantitatively, applying the A-EDT algorithm together with the meme dynamics features produces the following results:

- $\tau$ = 12hr, accuracy = 83.5%, most predictive features: 1.) community dispersion, 2.) #k-core blogs, 3.) #posts.
- $\tau$ = 24hr, accuracy = 91.5%, most predictive features: 1.) community dispersion, 2.) post rate, 3.) #posts.
- $\tau$ = 48hr, accuracy = 92.8%, most predictive features: 1.) community dispersion, 2.) post rate, 3.) #posts.
- $\tau$ = 120hr, accuracy = 97.5%, most predictive features: 1.) community dispersion, 2.) #posts, 3.) #ES blogs.

These results show that useful predictions can be obtained *within the first twelve hours* after a meme is detected (this corresponds to 0.5% of the average meme lifespan), and that accurate prediction is possible after about a day or two. Not also that, as has been found with other social dynamics phenomena [e.g., 11,12], dynamics features appear to be more predictive than "intrinsics", at least for the features employed here. These prediction results are summarized in Figure 7. In particular, the plot at the top depicts the prediction accuracy obtainable using just the most predictive intrinsic feature (ANEW arousal, red), and just the most predictive dynamics feature (community dispersion($\tau$), blue), as a function of the early time series available $\tau$. The middle plot shows the prediction accuracy provided by the full A-EDT algorithm (i.e., employing all the features) as a function of early time series available. The table at the bottom of Figure 7 summarizes these results.

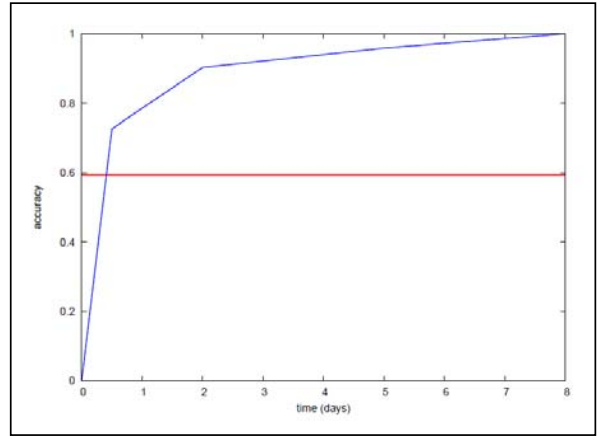

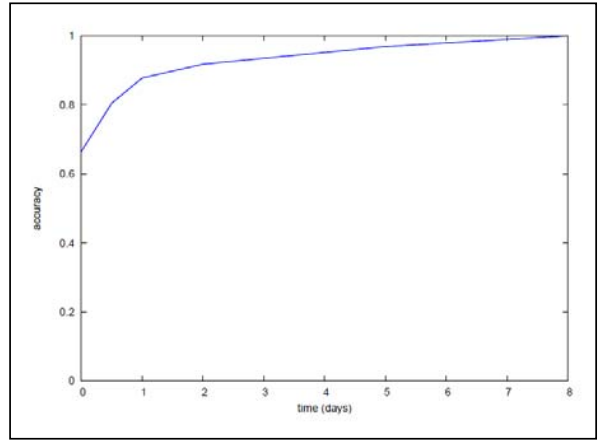

**Summary of Prediction Results**

| $\tau$ (% lifespan) | accuracy | ranked predictive features |
|---|---|---|
| 0 hr (0%) | 66.5% | 1.) arousal, 2.) sent., 3.) happiness. |
| 12 hr (0.5%) | 83.5% | 1.) comms., 2.) k-core, 3.) #posts. |
| 24 hr (1.0%) | 91.5% | 1.) comms., 2.) post rate, 3.) #posts. |
| 48 hr (1.9%) | 92.8% | 1.) comms., 2.) post rate, 3.) #posts. |
| 120 hr (4.8%) | 97.5% | 1.) comm., 2.) #posts, 3.) #ES blogs. |

**Figure 7.** Sample meme prediction results. Top plot shows the prediction accuracy obtainable when using the most predictive intrinsic feature alone (ANEW arousal, red), and most predictive dynamics feature alone (community dispersion($\tau$), blue), as a function of the early time series available. Middle plot depicts the prediction accuracy provided by the A-EDT algorithm as a function of the early time series available. Table at bottom summarizes the results, reporting the prediction accuracy (in %) and three most predictive features as a function of early time series available (in hours and as % of mean meme lifespan).

## Acknowledgements

This work was supported by the U.S. Department of Defense and the Laboratory Directed Research and Development program at Sandia National Laboratories.